\documentclass[a4paper,11pt]{article}

\usepackage{jcappub}
\usepackage{amsfonts}
\usepackage{amsmath}
\usepackage{url}
\usepackage{epstopdf}

\title{Is the continuous matter creation cosmology an alternative to $\Lambda$CDM?}

\author[a]{J. C. Fabris,}
\affiliation[a]{Departamento de F\'isica, Universidade Federal do Esp\'irito Santo, Vit\'oria, ES, Brazil}
\author[a,b]{J. A. de Freitas Pacheco,}
\affiliation[b]{Universit\'e de Nice-Sophia Antipolis, Observatoire de la C\^ote d'Azur, Laboratoire Lagrange, Nice Cedex 4, France}
\author[a]{and O. F. Piattella}

\emailAdd{fabris@pq.cnpq.br}
\emailAdd{pacheco@oca.eu}
\emailAdd{oliver.piattella@pq.cnpq.br}

\abstract{The matter creation cosmology is revisited, including the evolution of baryons and dark matter particles. The creation process affects only dark matter and not baryons. The dynamics of the $\Lambda$CDM model can be reproduced only if two conditions are satisfied: 1) the entropy density production rate and the particle density variation rate are equal and 2) the (negative) pressure associated to the creation process is constant. However, the matter creation model predicts a present dark matter-to-baryon ratio much larger than that observed in massive X-ray clusters of galaxies, representing a potential difficulty for the model. In the linear regime, a fully relativistic treatment indicates that baryons are not affected by the creation process but this is not the case for dark matter. Both components evolve together at early phases but lately the dark matter density contrast decreases since the background tends to a constant value. This behaviour produces a negative growth factor, in disagreement with observations, being a further problem for this cosmology.}


\arxivnumber{}

\begin{document}

\maketitle
\flushbottom

\section{Introduction}
\label{intro}

Presently, the standard cosmological model ($\Lambda$CDM) includes a constant term in Einstein equations and assumes that the universe is constituted, besides baryons, photons and neutrinos, by a dominant weakly interacting component of unknown nature dubbed dark matter. This model gives the best representation of different independent data sets as distances to type Ia supernovae, the angular power spectrum of the cosmic microwave background (CMB) or the angular distance scale of the baryon acoustic oscillations (BAO) (see, for instance, \cite{kowal08,tsuji04} and references therein).

Despite these successes, arguments against the inclusion of a cosmological term are often found in the literature. Among others, the so-called ``coincidence" problem and the dramatic difference between predicted and observed values of the cosmological constant, when interpreted as the vacuum energy density (however, see \cite{bianchi} for counter-arguments). Face to these hypothetical difficulties, alternative models able to explain the present accelerated phase of the cosmic expansion have been proposed in the past years (see \cite{tsuji04} for a short review). Among these models, we mention those in which a negative pressure, responsible for the present accelerated phase of the cosmic expansion, is associated to a process of particle creation \cite{zeldovich,calvao,zimdahl94,lima99,pavon,cardenas,lima09}. 

In \cite{roany} cosmological models based on creation of cold dark matter particles (hereafter simply CCDM) and on open thermodynamic systems \cite{prigogine} were reviewed and, in particular, the case in which the entropy density production rate is equal to the particle density variation rate. Models satisfying this condition include practically all CCDM scenarios found in the literature. The derived background dynamics mimics that of the $\Lambda$CDM cosmology only if the (negative) pressure related to the creation process is constant. However, in the linear regime the growth of density perturbations differs from the standard model because, after attaining a maximum amplitude that depends on the adopted creation rate, the density contrast decreases \cite{roany}. This is certainly a difficulty for this class of cosmology but these results were questioned by \cite{lima10}, who criticised the Newtonian approach adopted by \cite{roany} in their analysis of linear perturbations.

Since CCDM models are still being investigated as a viable alternative to the $\Lambda$CDM cosmology (see, for instance, \cite{ioav,pan,berezin,Chakraborty:2014oya}), we report here a new investigation that differs from previous analyses in two main aspects. Firstly, we consider the evolution of both dark matter and baryons either in the background or in the linear regime. Secondly, a fully relativistic analysis of the linear perturbations of both components is performed. As found by previous authors, we will show that the background dynamics of the CCDM cosmology is formally identical to the $\Lambda$CDM model and, consequently, satisfies the same tests mentioned above. However, in the CCDM model the density ratio between dark matter particles and baryons varies and presently, as we shall see, the theory predicts a value of about 19.5, which is considerably higher than that derived from the analysis of massive X-ray clusters. We will show that in the linear regime the growth of baryons is not affected by the particle creation process but that of dark matter is, as it was pointed out by \cite{roany}. These aspects represent potential problems to be surpassed by the CCDM cosmology.  This paper is organised as follows: in Section~\ref{Sec2} the main properties of the CCDM model are reviewed, in Section~\ref{Sec3} the relativistic linear equations are derived and finally, in Section~\ref{Sec4} the main conclusions are given.  

\section{The CCDM model}\label{Sec2}

In the CCDM cosmology, non-relativistic dark matter particles are assumed to be produced continuously and are supposed to act like a fluid. According to \cite{prigogine}, consider that the universe is an open thermodynamic system in which particles are created at the expense of the gravitational field. Then, let $\mu$ be the chemical potential associated to the variation of the number of particles. In this case, the Euler relations can be written as
\begin{equation}
\label{eum}
\mu n = \mathfrak{h} -Ts\;,
\end{equation}
and
\begin{equation}
Tds = d\rho - \mu dn\;.
\label{edois}
\end{equation}
In the equations above, $n$ is the dark matter particle density, $s$ and $\rho$ are respectively the entropy and the energy densities, $\mathfrak{h}=P+\rho$ is the enthalpy density and $T$ is the temperature. From these equations one obtains trivially
\begin{equation}
\label{etres}
\left(\frac{d\rho}{dt}-\frac{\mathfrak{h}}{n}\frac{dn}{dt}\right)=sT\left(\frac{1}{s}\frac{ds}{dt}-
\frac{1}{n}\frac{dn}{dt}\right)\;,
\end{equation}
which is the Gibbs relation for the dark matter fluid. As already mentioned, a common assumption in most CCDM models is that the right side of eq.~\eqref{etres} be zero. This is equivalent to say that the entropy production and the particle variation rates are equal. In other words, the entropy per particle $s/n$ remains constant during the expansion process. It should be emphasized that although this last condition be also verified in the $\Lambda$CDM model, there is an important difference with respect to the CCDM cosmology: in the former the expansion is adiabatic while in the later entropy is produced as a consequence of the particle creation process. In this case, under the assumption of a constant $s/n$ ratio eq.~\eqref{etres} becomes simply
\begin{equation}
\label{equatro}
\frac{d\rho}{dt}=\frac{\mathfrak{h}}{n}\frac{dn}{dt}\;.
\end{equation}
Following \cite{roany}, let us introduce the stress-energy tensor for the dark matter fluid as
\begin{equation}
\label{ecinco}
T_{ik}=(\rho + \Pi)u_iu_k - \Pi g_{ik}\;,
\end{equation}
where $\Pi = P(\rho)+P_{\rm c}$ is the effective pressure acting on the fluid, with the first term representing the pressure due to kinetic motions and interactions between particles while the second represents the effective pressure associated to the process of particle creation. Assuming a flat Friedmann-Robertson-Walker (FRW) metric, i.e.,
\begin{equation}
\label{eseis}
ds^2 = -dt^2 + a^2(t)\left(dr^2 + r^2d\Omega^2\right)\;,
\end{equation}
and using the conservation equation $T^k{}_{0;k}=0$, one obtains
\begin{equation}
\label{esete}
\frac{d\rho}{dt}+3H(\mathfrak{h}+P_{\rm c})=0\;,
\end{equation}
where, as usual, $H={\dot a}/a$ is the Hubble parameter. Combining now eqs.~\eqref{equatro} and \eqref{esete}, an explicit expression for the pressure associated to the particle creation process can be obtained, i.e., 
\begin{equation}
\label{eoito}
P_{\rm c} = -\frac{\mathfrak{h}}{3H}\left(3H+\frac{1}{n}\frac{dn}{dt}\right)\;.
\end{equation}
Note that if ${\dot n}>0$ and the background is expanding ($H>0$), the pressure $P_{\rm c}$ is negative. Moreover, if the created particles are ``cold" (non-relativistic) or, in other words, their kinetic energy and pressure can be neglected, their enthalpy is simply 
$\mathfrak{h}=\rho=mnc^2$, where $m$ is the mass of the dark matter particles. In this case, eq.~\eqref{eoito} can be rewritten as
\begin{equation}
3HP_{\rm c} = -3Hnmc^2-mc^2\frac{dn}{dt}\;.
\end{equation}
Multiplying both sides of this equation by $a^3$, one obtains easily
\begin{equation}
\label{enove}
P_{\rm c}\frac{da^3}{dt}+\frac{d(nmc^2a^3)}{dt}=0\;.
\end{equation} 
The equation above permits a simple interpretation of the creation pressure $P_{\rm c}$. The work done by this stress to expand a unit comoving volume of the universe is equal to the rate of energy in the form of new particles appearing in the same comoving volume. A past study by \cite{mccrea} on matter creation in the context of general relativity led to the same interpretation.

\subsection{The Boltzmann-Einstein approach}

In the previous Section, the basic equation defining $P_{\rm c}$ was derived using a thermodynamic approach and assuming an open system. Here, for the sake of completeness, the Boltzmann equation in a curved space is used to re-derive the equations of the model.

With the metric given by eq.~\eqref{eseis}, the Boltzmann equation for the dark matter fluid is \cite{oliver}
\begin{equation}
\label{boltzmann}
\frac{\partial f}{\partial t}-Hp\frac{\partial f}{\partial p}=C[f]\;.
\end{equation}
In the equation above, using the same notation as in \cite{oliver}, $p$ is the modulus of the 3-momentum and $f$ is the one-particle distribution function satisfying the condition
\begin{equation}
\label{density}
n = \int f(p)d^3p\;.
\end{equation}
The right side of eq.~\eqref{boltzmann} is not a truly collisional term but a source term intending to represent phenomenologically the creation process. A simple expression for this term can be derived in the following way. First, we integrate over momenta eq.~\eqref{boltzmann} to obtain
\begin{equation}
\label{firstmomentum}
{\dot n}+3Hn = \int C[f]d^3p\;.
\end{equation}
Comparing the equation above with eq.~\eqref{esete} and recalling that $h=\rho=nmc^2$, one obtains
\begin{equation}
\label{collisionterm}
\int C[f]d^3p = -\frac{3HP_{\rm c}}{mc^2}\;.
\end{equation}  
In a further step, assume the following ansatz for the distribution function, $f(p,a)=f_0(pa)g(a)$, where $f_0(pa)$ is a solution of eq.~\eqref{boltzmann} without the source or ``collisional" term (see \cite{oliver}). Replacing this into eq.~\eqref{boltzmann}, one obtains
\begin{equation}
\label{collisionb}
f_0\frac{dg(a)}{da}Ha = C[f]\;.
\end{equation}
Substitute the equation above into eq.~\eqref{collisionterm} to get, after some straightforward calculations, 
\begin{equation}
\label{enoveb}
\frac{dg(a)}{da}=-\frac{3P_{\rm c}}{\rho(a)}\frac{g(a)}{a}\;.
\end{equation}
The solution of eq.~\eqref{esete} gives the evolution of the dark matter energy density and is given by eq.~\eqref{equatorze} for the case in which the creation pressure is constant, an assumption also made here. Replacing the expression for $\rho(a)$ into eq.~\eqref{enoveb} gives
\begin{equation}
\label{enovec}
\frac{dg(a)}{da}=\frac{3a^2}{a^3+\beta}g(a)\;,
\end{equation}
where the parameter $\beta$ was defined by
\begin{equation}
\label{beta}
\beta + 1=-\frac{\rho_{\rm dm0}}{P_{\rm c}}\;,
\end{equation}
with $\rho_{\rm dm0}$ being the present dark matter energy density. Integration of eq.~\eqref{enovec} is trivial and one gets $g(a)=K(\beta+a^3)$, where $K$ is an integration constant. Now, substitute this result into eq.~\eqref{collisionb} to obtain 
\begin{equation}
\label{source1}
C[f]=3KHa^3f_0\;.
\end{equation}
Using these results, the distribution function that satisfies eq.~\eqref{boltzmann} with the collisional term above is $f=Kf_0(\beta+a^3)$. Combining this with eq.~\eqref{source1} permits to recast the collisional term as
\begin{equation}
\label{source}
C[f]=\frac{3Ha^3}{\beta + a^3}f\;.
\end{equation}
Note that when $P_{\rm c}$ goes to zero from the negative side, $\beta$ goes to infinity and the collisional term goes to zero as wished, since no particles are being created. Using the same reasoning, the integration constant may be put equal to $K=1/\beta$ since in the limit of zero creation pressure, the distribution function must be essentially $f_0$, the solution of the Vlasov-Einstein equation.

In a further step, replace the derived expression for the source term into eq.~\eqref{boltzmann}. Then multiply both sides of the equation by the particle energy $E$ and integrate over the 3-momentum to obtain
\begin{equation}
\label{energy1}
{\dot\rho}+3H(\rho + P)=\frac{3Ha^3}{\beta + a^3}\rho\;,
\end{equation}
where the energy density and the pressure were defined as usually, i.e.,
\begin{equation}
\rho = \int Efd^3p\;,
\end{equation}
and
\begin{equation}
\label{pressure}
P = \int\frac{p^2}{3E}fd^3p\;.
\end{equation}
Moving the right side of eq.~\eqref{energy1} to the left side, it is easy to verify that this is equivalent to add to the kinetic pressure a term equal to
\begin{equation}
P_{\rm c} = -\frac{\rho a^3}{\beta + a^3}\;.
\end{equation}
Since a constant creation pressure was assumed, the equation above is, as expected, the same as that resulting from the integration of eq.~\eqref{esete}, indicating consistency between both approaches.

It is interesting to mention that the creation process affects the kinetic pressure of dark matter particles. This can be seen integrating eq.~\eqref{pressure} with a new variable $x = pa$. In this case, one obtains
\begin{equation}
P=\frac{4\pi}{3}\frac{\beta + a^3}{\beta a^4}\int dxf_0(x)\frac{x^4}{\sqrt{x^2+m^2a^2}}\;.
\end{equation}
Since the created particles are non-relativistic, they satisfy the condition $m^2a^2 \gg x^2$. In this case, the above integral can be simplified and one obtains for the pressure
\begin{equation}
P=\frac{4\pi}{3m}\frac{\beta + a^3}{\beta a^5}\int x^4f_0(x)dx\;.
\end{equation}
When particles are not being created ($\beta \rightarrow \infty$), the kinetic pressure decays as $P\propto a^{-5}$, a well known result. However, when $\beta$ is finite, at high redshifts the pressure decays as expected but at redshift of the order of $1+z\sim 1/\beta$, the logarithmic slope $(d\log~P/d\log~a)$ is higher than -5, indicating effects of the creation process. 

\subsection{Dynamics of the CCDM model}

In the present study the contribution of both baryons and dark matter to the dynamics of the universe is taken into account. However, the continuous particle creation process affects only the dark matter component, since baryons are supposed to be conserved. As it was shown in \cite{roany}, the CCDM model can mimic the $\Lambda$CDM cosmology only if a {\it constant} creation pressure is assumed, which here will be denoted as $P_{\rm c} = -\lambda$, where $\lambda$ is a positive constant with the dimension of an energy density. Assuming further that the thermodynamic pressure of baryons and dark matter could be neglected, the stress-energy tensor for dark matter and baryons can be recast as
\begin{equation}
\label{edez}
T_{ik}({\rm dm})=\left(\rho_{\rm dm}-\lambda\right)u_iu_k + \lambda g_{ik}\;,
\end{equation}
and
\begin{equation}
\label{eonze}
T_{ik}({\rm b})=\rho_{\rm b}u_iu_k\;,
\end{equation}
The subscripts ``dm" and ``b" refer respectively to dark matter and baryons. The conservation equations $T^k{}_{0;k}(\rm dm)=0$ and $T^k{}_{0;k}(\rm b)=0$ give respectively
\begin{equation}
\label{edoze}
\frac{d\rho_{\rm dm}}{dt}+3H\rho_{\rm dm} = 3H\lambda\;,
\end{equation}
and
\begin{equation}
\label{etreze}
\frac{d\rho_{\rm b}}{dt}+3H\rho_{\rm b} = 0\;.
\end{equation}
These equations are easily integrated in terms of the scale factor $a$ and one obtains
\begin{equation}
\label{equatorze}
\rho_{\rm dm} = \lambda + \frac{(\rho_{\rm dm0}-\lambda)}{a^3}\;,
\end{equation}
and
\begin{equation}
\label{equinze}
\rho_{\rm b} = \frac{\rho_{b0}}{a^3}\;.
\end{equation}
The constants $\rho_{\rm dm0}$ and $\rho_{b0}$ denote the present ($a=1$) values of the dark matter and the baryon energy densities.

Using eqs.~\eqref{equatorze} and \eqref{equinze}, the Hubble equation can be written as
\begin{equation}
\label{edezesseis}
H^2 = \frac{8\pi G}{3c^2}\left[\lambda+\frac{(\rho_{\rm dm0}+\rho_{\rm b0}-\lambda)}{a^3}\right]\;.
\end{equation}
The first point to be noticed in this equation is that the present value of the Hubble parameter is fixed only by the present total matter content $\rho_{\rm t0} = \rho_{\rm dm0} + \rho_{\rm b0}$ of the universe as in the Einstein-de Sitter model, namely
\begin{equation}
\label{hubble}
H^2_0 = \frac{8\pi G}{3c^3}\rho_{\rm t0}\;.
\end{equation}
The second point is that eq.~\eqref{edezesseis} can be written in a form equivalent to the $\Lambda$CDM model, i.e., $H^2=H^2_0(\Omega_{\rm v}+\Omega_{\rm m0}a^{-3})$, if the following identifications are made
\begin{equation}
\label{omega}
\Omega_{\rm v} = \frac{\lambda}{\rho_{\rm t0}}\;,
\end{equation}
and
\begin{equation}
\label{omegam}
\Omega_{\rm m0} = \frac{(\rho_{\rm dm0}+\rho_{\rm b0}-\lambda)}{\rho_{\rm t0}}\;.
\end{equation}
Since the Hubble equation in the CCDM model is formally identical to that of the $\Lambda$CDM cosmology, one should expect that data on supernova distances, BAO angular distances, variation of the Hubble parameter with $z$ will be equally explained by both cosmological models. This identification led to different authors to conclude that CCDM and $\Lambda$CDM models are indistinguishable at the background level. However there are some subtleties in this analysis. From the equations above, the ratio between dark matter and baryons energy densities is
\begin{equation}
\label{edezesete}
\frac{\rho_{\rm dm}}{\rho_{\rm b}}=\frac{\left[(1-\Omega_{\rm m0})a^3+(\Omega_{\rm m0}-\Omega_{\rm b0}\right]}{\Omega_{\rm b0}}\;.
\end{equation}
The cosmological data from Planck \cite{planck} gives for the total matter density parameter $\Omega_{\rm m0}=0.315$ whereas that of the baryons is $\Omega_{\rm b0}=0.04872$. Using these values, eq.~\eqref{edezesete} at the baryon-radiation decoupling or at still higher $z$ gives a ratio $\rho_{\rm dm}/\rho_{\rm b}$ = 5.46. This is essentially the Planck value for the $\Lambda$CDM model, implying that the CCDM cosmology predicts the same amplitude for the peaks of the angular power spectrum of the CMB and again both models agree. However, since dark matter particles are being produced continuously, the present value of the ratio between dark matter and baryon energy densities predicted by the CCDM model is $\rho_{\rm dm}/\rho_{\rm b}$ = 19.5. Observations of clusters of galaxies in the mass range $6\times 10^{13} - 1\times 10^{15} M_{\odot}$ reported by \cite{gonzalez} indicate a nearly constant value for this ratio and equal to $\rho_{\rm dm}/\rho_{\rm b}$ = 6.52. The $\rho_{\rm dm}/\rho_{\rm b}$ ratio derived from massive clusters is about 20\% higher than the ``cosmic" value derived from CMB data and the authors of this study suggest that a small baryon deficiency could be present, consequence of different processes operating in the formation of these large structures as, for instance, the feedback of AGNs. Turning the argument, this value also imposes some constraints on the possible amount of dark matter created after decoupling, which is considerably smaller than the growth by a factor of about 3.6 predicted by the CCDM model. This discrepancy represents a potential problem for the CCDM model already at the background level and may represent a test able to distinguish both cosmologies. 

However, the authors of reference \cite{ioav} have a different interpretation for the evolution of the baryon-to-dark matter ratio. They assume that the dark matter energy density is the sum of two distinct terms: the first, denoted by $\rho_{\rm cre}$ corresponds to particles that are being created continuously, forming a uniform background. According to them, these particles do not participate in the formation process of structures; the second term, denoted by $\rho_{\rm con}$, corresponds to particles formed in a short timescale at very high redshifts, which are ``conserved" and are able to cluster. In eq.~\eqref{equatorze}, they made the identifications $\rho_{\rm cre}=\lambda$ and $\rho_{\rm con}= (\rho_{\rm dm0}-\lambda)/a^3$. Clearly, if only ``conserved" particles participate in the formation of structures, the aforementioned problem with the baryon-to-dark matter ratio disappear. Such an interpretation can be criticised by the following reasons. The first concerns the identification of $\rho_{\rm cre}$ with $\lambda$. Since the density of this component is constant, this implies that newly created particles are permanently in a steady state or, in other words, the rate at which particles are being created compensates exactly the losses due to the expansion of the universe, requiring fine tuning that represents a very special situation. The second aspect concerns the clustering of the ``conserved" term only. This is based on the fact that the perturbation of the dark matter energy density is
\begin{equation}
\delta\rho_{\rm dm}=\delta\left(\rho_{\rm cre}+\rho_{\rm con}\right)=\delta\rho_{\rm con}\;,
\end{equation}
since $\delta\rho_{\rm cre}=0$, because $\rho_{\rm cre}$ is a constant. In this case, only the term $\delta\rho_{\rm con}$ contributes to the Poisson equation or to the gravitational potential, leading to the idea that particles created continuously do not give any contribution to the potential and thus, they do not cluster. In other words, the newly created particles would not feel the gravitational forces, representing a unlikely situation since all dark matter particles are expected to be indistinguishable. In the general case, using the same components as defined in \cite{ioav}, insert the total dark matter energy density $\rho_{\rm dm} = \rho_{\rm cre}+\rho_{\rm con}$ into eq.~\eqref{edoze} that can be split in two equations: one including the source term, which is obeyed by the component $\rho_{\rm cre}$ and another, without the source term, which is satisfied by the component $\rho_{\rm con}$. Under these conditions, the general solution for the created component is
\begin{equation}
\rho_{\rm cre}=\lambda + \frac{K_1}{a^3}\;.
\end{equation}
The integration constant $K_1$ can be estimated by requiring that when $\lambda \rightarrow 0$ the created component must disappear. Then, $K_1=\lambda$. On the other side, the general solution for the conserved component is
\begin{equation}
\rho_{\rm con}=\frac{K_2}{a^3}\;.
\end{equation}
The integration constant $K_2$ can be derived by imposing that at $a=1$ (present time) the dark matter energy density be equal to $\rho_{\rm dm0}$. Consequently, one obtains $K_2=(\rho_{\rm dm0}-2\lambda$) and hence the total dark matter energy density is recovered by adding both components, i.e.,
\begin{equation}
\rho_{\rm dm}=\left[\lambda + \frac{\lambda}{a^3}\right]+\left[\frac{(\rho_{\rm dm0}-2\lambda)}
{a^3}\right]=\lambda+\frac{(\rho_{\rm dm0}-\lambda)}{a^3}\;.
\end{equation} 
By such a procedure eq.~\eqref{equatorze} is recovered and it becomes clear that the ``conserved" term includes both components in contradiction with the hypothesis raised by \cite{ioav}.

\section{Linear perturbations}\label{Sec3}

In the present investigation, the evolution of density perturbations either of the baryonic or of the dark matter fluids will be considered using a fully relativistic approach. We follow here the treatment and the notation as in \cite{ma}, adopting a synchronous gauge. In this case the perturbed metric can be written as (the convention $c=1$ is adopted, excepting cases when recovering all constants is necessary)
\begin{equation}
ds^2=-dt^2+a^2(t)\left(\delta_{ik}+h_{ik}\right)dx^idx^k\;.
\end{equation}
Combining eqs. 20a and 20c given in reference \cite{ma} but including both dark matter and baryons and taking the derivatives with respect to the cosmic time and not with respect to the conformal time one obtains
\begin{equation}
\label{pum}
{\ddot h} + 2H{\dot h} = -8\pi G\left(\rho_{\rm dm}\delta_{\rm dm}+\rho_{\rm b}\delta_{\rm b}\right)\;,
\end{equation}
where $h$ is the trace of the metric perturbation tensor $h_{ik}$ and, as usually, the density contrast for baryons $\delta_{\rm b} = \delta\rho_{\rm b}/\rho_{\rm b}$ and for dark matter $\delta_{\rm dm}=\delta\rho_{\rm dm}/\rho_{\rm dm}$ was introduced.

\begin{figure}
\center
\includegraphics[width=0.5\columnwidth]{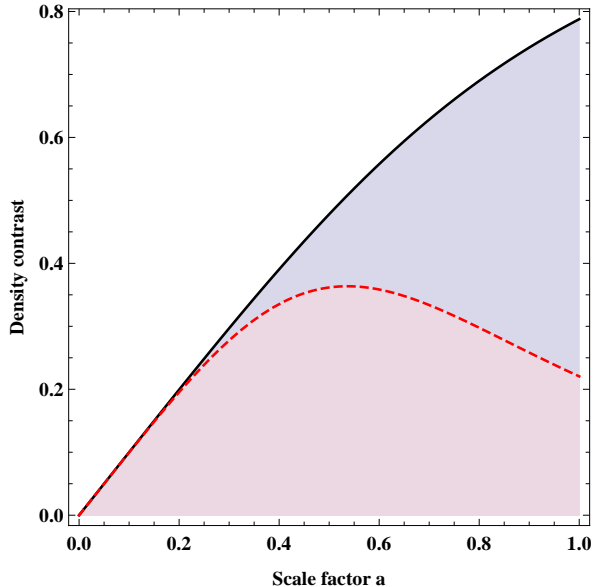}
\caption{Evolution of the density contrast for baryons (upper curve) and dark matter (lower curve as a function of the scale factor $a$.}
\label{fig:1} 
\end{figure}

Perturbations of the conservation equations are given by eqs. 28 and 29 of reference \cite{ma}. Here they are explicitly given for the two fluid components under the following assumptions: zero sound velocity for both fluids and the parameter $w=P/\rho$ being zero for baryons and equal to $w=-\lambda/\rho_{\rm dm}$ for dark matter. These approximations imply that the perturbations are adiabatic.
However, since there is entropy production in the matter creation process,
non-adiabatic perturbations are also possible and they will be analyzed in a
future work. Taking again the derivatives with respect to the cosmic time one obtains
\begin{equation}
\label{pdois}
{\dot\delta_{\rm b}}=-\frac{1}{2}{\dot h}-\frac{\theta_{\rm b}}{a}\;,
\end{equation}
and
\begin{equation}
\label{ptres}
{\dot\delta_{\rm dm}}=-(1+w)\left(\frac{{\dot h}}{2}+\frac{\theta_{\rm dm}}{a}\right)+3Hw\delta_{\rm dm}\;,
\end{equation}
where $\theta = \delta u^i{}_{;i}$ is the divergence of the peculiar velocity, obeying the equations below, derived also from eqs. 28 and 29 given in reference \cite{ma} and under the assumptions mentioned above,
\begin{equation}
\label{pquatro}
{\dot\theta_{\rm b}}=-H\theta_{\rm b}\;,
\end{equation}
and
\begin{equation}
\label{pcincoa}
{\dot\theta_{\rm dm}}=-(1-3w)H\theta_{\rm dm}-\frac{{\dot w}}{1+w}\theta_{\rm dm}\;.
\end{equation}
Performing the time derivative of the relation $w=-\lambda/\rho_{\rm dm}$ and using eq.~\eqref{edoze} one obtains
\begin{equation}
\label{evinte}
{\dot w}=3Hw(1+w)\;.
\end{equation}
Replacing this result into eq.~\eqref{pcincoa}, one obtains simply
\begin{equation}
\label{pcincob}
{\dot\theta_{\rm dm}}=-H\theta_{\rm dm}\;.
\end{equation}
The integration of eqs.~\eqref{pquatro} and \eqref{pcincob} is trivial and one obtains $\theta_{\rm b} \propto 1/a$ and $\theta_{\rm dm} \propto 1/a$. These are decaying solutions that justify to neglect the $\theta$ terms in eqs.~\eqref{pdois} and \eqref{ptres}. Under
this condition, eq.~\eqref{pdois} simplifies to
\begin{equation}
\label{pdoisb}
{\dot\delta_{\rm b}}=-\frac{1}{2}{\dot h}\;.
\end{equation}
Replacing this equation and its time derivative into eq.~\eqref{pum} results
\begin{equation}
\label{pumb}
{\ddot\delta_{\rm b}}+2H{\dot\delta_{\rm b}}-4\pi G(\rho_{\rm b}\delta_{\rm b}+\rho_{\rm dm}\delta_{\rm dm})=0\;.
\end{equation}
Neglecting the $\theta$ term also in eq.~\eqref{ptres} and using eq.~\eqref{pdoisb} results
\begin{equation}
\label{ptresb}
{\dot\delta_{\rm dm}}=(1+w){\dot\delta_{\rm b}}+3Hw\delta_{\rm dm}\;.
\end{equation}
A further step in the solution of this equation is to introduce the reduced dark matter density contrast by the relation
\begin{equation}
\label{pseis}
{\tilde\delta_{\rm dm}}=\frac{\delta\rho_{\rm dm}}{(\rho_{\rm dm}-\lambda)}=
\frac{\delta_{\rm dm}}{1+w}\;.
\end{equation}
Taking the time derivative of this equation and using eq.~\eqref{evinte} one obtains
\begin{equation}
\label{psete}
\frac{d{\delta_{\rm dm}}}{dt}=(1+w)\frac{d{\tilde\delta_{\rm dm}}}{dt}+3Hw(1+w)
{\tilde\delta_{\rm dm}}\;.
\end{equation}
Substituting eq.~\eqref{psete} into eq.~\eqref{ptresb} gives simply $d{\tilde\delta_{\rm dm}}/dt=d\delta_{\rm b}/dt$ or ${\tilde\delta_{\rm dm}}=\delta_{\rm b}+ \rm constant$. Taking the integration constant equal to zero and making use of the definition of the reduced density contrast, one obtains finally
\begin{equation}
\label{dmcontrast}
\delta_{\rm dm}=(1+w)\delta_{\rm b}\;.
\end{equation}
Replacing the equation above into eq.~\eqref{pumb} one gets for the evolution of the baryon density contrast
\begin{equation}
\label{pumc}
{\ddot\delta_{\rm b}}+2H{\dot\delta_{\rm b}}-4\pi G\left[\rho_{\rm b}+(1+w)\rho_{\rm dm}\right]\delta_{\rm b}=0\;.
\end{equation}
Making use of eqs.~\eqref{hubble}, \eqref{edezesete} and after some algebra, the equation above can be recast as
\begin{equation}
\label{pumd}
{\ddot\delta_{\rm b}}+2H{\dot\delta_{\rm b}}-\frac{3}{2}H_0^2\Omega_{\rm m0}a^{-3}\delta_{\rm b}=0\;.
\end{equation}
This equation for the evolution of the baryon density contrast is formally the same as that for the $\Lambda$CDM model, indicating that baryons are not directly affected by the matter creation process. However, in the CCDM model, the evolution of the dark matter density contrast and that of the baryons are not the same as it occurs in the $\Lambda$CDM cosmology. They are related by eq.~\eqref{dmcontrast} and since the parameter $w$ is not a constant, the evolution of both components is not proportional to each other. Figure \ref{fig:1} shows the evolution of the density contrast for both fluids, using the cosmological parameters given in reference \cite{planck}. The linear growth of baryons coincides with that of the $\Lambda$CDM model but the dark matter fluid displays a different behaviour. After reaching a maximum around $z\sim$ 1, the amplitude of the density contrast decreases, in agreement with the conclusions by \cite{roany}, who have performed a Newtonian analysis and with those of \cite{ioav}, who have performed an one-fluid investigation based either on the Neo-Newtonian approximation or on a relativistic approach.
How to explain the behaviour of the dark matter density contrast? The evolution of the background dark matter density is given by eq.~\eqref{equatorze}. This equation indicates that at high $z$ (or $a \ll 1$) the dark matter background evolves as the baryon background, i.e., inversely proportional to the cube of the scale parameter $a$. Using the spherical model as a guide, in the early phases of the linear regime the background  ``expands" faster than the density perturbation, leading to an increase of density contrast for both fluids as it can be seen in fig.~\ref{fig:1}. However, as a consequence of the particle creation process, in late phases the dark matter background varies slower than the baryon background (not affected in the creation process), tending to a constant value fixed by the parameter $\lambda$. Such a modification in the background expansion rate produces a decrease in the amplitude
of the dark matter density contrast. A further argument in favour of this explanation is provided by eq.~\eqref{dmcontrast}. Using eqs.~\eqref{omega} and \eqref{omegam}, the eq.~\eqref{dmcontrast} can be recast as 
\begin{equation}
\delta_{\rm dm}=\left[1+\frac{\Omega_{\rm v}a^3}{(\Omega_{\rm m0}-\Omega_{\rm b0})}\right]^{-1}\delta_{\rm b}\;.
\end{equation}
This equation shows clearly that when $a\ll 1$, the contrast of both components is the same but as soon as the particle creation term becomes relevant, the dark matter density contrast decreases with respect to that of the baryon component, since now the background of both fluids evolves differently as explained previously.
\begin{figure}
\center
\includegraphics[width=0.5\columnwidth]{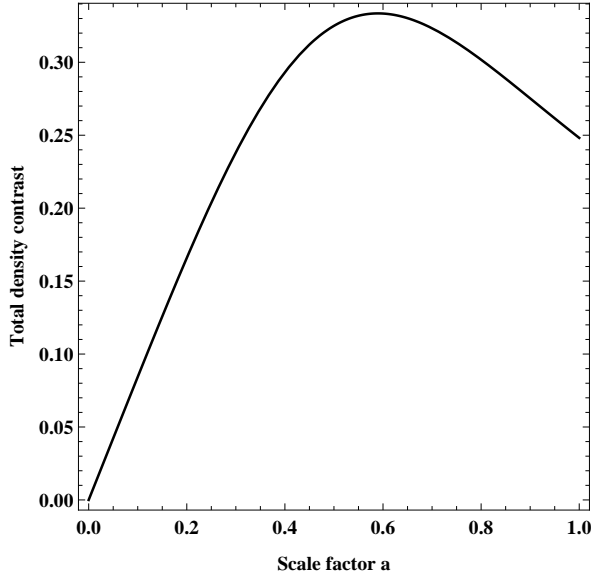}
\caption{Evolution of the total matter density contrast as a function of the scale factor $a$.}
\label{fig:2} 
\end{figure}
In figure \ref{fig:2} is shown the evolution of the total matter density contrast defined by
\begin{equation}
\delta_{\rm m} =\frac{\rho_{\rm b}}{(\rho_{\rm b}+\rho_{\rm dm})}\delta_{\rm b}+\frac{\rho_{\rm dm}}{(\rho_{\rm b}+\rho_{\rm dm})}\delta_{\rm dm}\;.
\end{equation}  
Since the dark matter component is dominant, the behaviour of the total matter density contrast is similar to that of dark matter.
\begin{figure}
\center
\includegraphics[width=0.5\columnwidth]{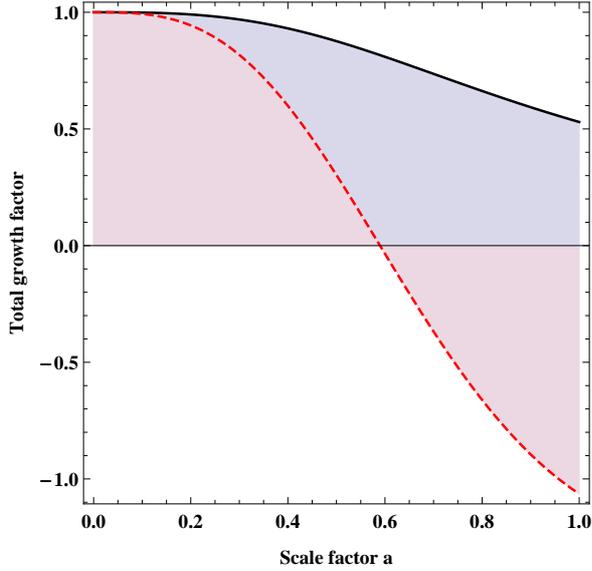}
\caption{Evolution of the growth factor for the $\Lambda$CDM model (upper curve) and for the CCDM model (lower curve) as a function of the scale parameter $a$.}
\label{fig:3}
\end{figure}
In figure \ref{fig:3}, the growth factor $f(z)=d\log\delta_{\rm m}/d\log a$ is shown for both the standard model and the CCDM model. Note that the decreasing amplitude branch of the density contrast in the CCDM model implies in {\it negative} values for growth factor, in disagreement with observational data given by \cite{hudson}, but that are 
consistent with predictions of the $\Lambda$CDM model. 

Different conclusions were reached by \cite{ioav} since they have defined a density contrast only for the component able to cluster. In this case, the evolution of the density contrast has a weak dependence on the parameter $\lambda$ and the growth factor is not inconsistent with data.

\section{Conclusions}\label{Sec4}

The main physical properties of the creation cold dark matter (CCDM) cosmology were revisited. When the thermodynamical approach by \cite{prigogine} is considered, the CCDM model is able to mimic the standard model if two main assumptions are made: the first is to consider that the entropy density production rate is equal to the particle variation rate and the second, is to hypothesise that the pressure associated to the creation process is constant. Generally, even if the creation pressure is not assumed to be constant, the first hypothesis and the conservation equation $T^k{}_{0;k}(\rm dm)=0$ lead to a simple interpretation of such a pressure: it corresponds to the work required to expand a unit comoving volume, which is equal to the energy of the (``cold") particles created in the same volume.

The solution of the Boltzmann-Einstein equation, including a pseudo-collisional term intending to represent the matter creation process, is consistent with the thermodynamical approach. The cosmic evolution of the kinetic or thermodynamic pressure of the newly created particles is affected by the creation process itself. At high redshifts, the kinetic pressure varies as $P\propto a^{-5}$, as expected for the standard model but at late phases the pressure decreases with a slower rate, consequence of the creation process.  

The resulting Hubble equation for the CCDM cosmology is formally identical to that of a flat $\Lambda$CDM model. Thus, both models are indistinguishable with respect to tests like the supernova distances, the angular distances of the baryon acoustic oscillations and the variation of the Hubble parameter with redshift. However, fixing the parameters of the Hubble equation in order to identify both models, an important difference between both cosmologies become evident. Since dark matter particles are being created continuously, the ratio between dark matter and baryon energy densities varies. At high redshift it agrees with Planck's data and no conflict with predicted and observed amplitudes of the angular power spectrum of the CMB exists. However, the dark matter-to-baryon ratio increases with decreasing redshift and presently is expected to be about 19.5. Such a high value disagrees with data on massive clusters of galaxies, which indicates values about 3 times smaller. This is certainly a difficulty for the CCDM model.

Previous analysis of the density contrast evolution in the linear regime using a Newtonian approximation has shown that there is a maximum amplitude that depends on the creation rate \cite{roany}. The present investigation adopts a fully relativistic approach and takes into account both baryons and dark matter particles. The present study indicates that baryons are not affected by the creation process and that the growth of perturbations in the linear regime coincides with that in the $\Lambda$CDM model. However, this is
not the case for dark matter particles. After reaching a maximum amplitude around $z\sim$1, the density contrast decreases as a consequence of the matter creation process that modifies the evolution of background, as explained in the previous section. The decreasing branch of the matter density contrast produces a negative growth rate of cosmic structures in contradiction with observational data. The difficulties pointed here raise doubts about the CCDM theory as a viable alternative to the standard model.

\acknowledgments

JAFP thanks the program ``Science without Borders" of the Brazilian Government for the financial support to this research and the Federal University of the Espírito Santo State (Brazil) for his hospitality. The authors thank CNPq and Fapes for support and are also grateful to Ioav Waga for discussions.

\bibliographystyle{JHEP}

\begin{thebibliography}{}
\bibitem{kowal08} M.~Kowalski {\it et al.}  [Supernova Cosmology Project Collaboration],
  ``Improved Cosmological Constraints from New, Old and Combined Supernova Datasets,''
  Astrophys.\ J.\  {\bf 686} (2008) 749
  [arXiv:0804.4142 [astro-ph]].
\bibitem{tsuji04} S.~Tsujikawa,
  ``Dark energy: investigation and modeling,''
  arXiv:1004.1493 [astro-ph.CO].
\bibitem{bianchi} E.~Bianchi and C.~Rovelli,
  ``Why all these prejudices against a constant?,''
  arXiv:1002.3966 [astro-ph.CO].
\bibitem{zeldovich} Y.~.B.~Zeldovich,
  ``Particle production in cosmology,''
  Pisma Zh.\ Eksp.\ Teor.\ Fiz.\  {\bf 12} (1970) 443.
\bibitem{calvao}  M.~O.~Calvao, J.~A.~S.~Lima and I.~Waga,
  ``On the thermodynamics of matter creation in cosmology,''
  Phys.\ Lett.\ A {\bf 162} (1992) 223.
\bibitem{zimdahl94} W.~Zimdahl and D.~Pavon,
  ``Cosmology with adiabatic matter creation,''
  Int.\ J.\ Mod.\ Phys.\ D {\bf 3} (1994) 327.
\bibitem{lima99} J.~A.~S.~Lima and J.~S.~Alcaniz,
  ``Flat FRW cosmologies with adiabatic matter creation: Kinematic tests,''
  Astron.\ Astrophys.\  {\bf 348} (1999) 1
  [astro-ph/9902337].
\bibitem{pavon} W.~Zimdahl, D.~J.~Schwarz, A.~B.~Balakin and D.~Pavon,
  ``Cosmic anti-friction and accelerated expansion,''
  Phys.\ Rev.\ D {\bf 64} (2001) 063501
  [astro-ph/0009353].
\bibitem{cardenas} V.~H.~Cardenas,
  ``Accelerated expansion and matter creation,''
  arXiv:0812.3865 [astro-ph].
\bibitem{lima09} J.~A.~S.~Lima, J.~F.~Jesus and F.~A.~Oliveira,
  ``CDM Accelerating Cosmology as an Alternative to LCDM model,''
  JCAP {\bf 1011} (2010) 027
  [arXiv:0911.5727 [astro-ph.CO]].
\bibitem{roany}  A.~de Roany and J.~A.~d.~F.~Pacheco,
  ``Continuous matter creation and the acceleration of the universe: The Growth of density fluctuations,''
  Gen.\ Rel.\ Grav.\  {\bf 43} (2011) 61
  [arXiv:1007.4546 [gr-qc]].
\bibitem{prigogine} I.~Prigogine, J.~Geheniau, E.~Gunzig and P.~Nardone,
  ``Thermodynamics and cosmology,''
  Gen.\ Rel.\ Grav.\  {\bf 21} (1989) 767.
\bibitem{lima10} J.~A.~S.~Lima, J.~F.~Jesus and F.~A.~Oliveira,
  ``Note on 'Continuous matter creation and the acceleration of the universe: the growth of density fluctuations',''
  Gen.\ Rel.\ Grav.\  {\bf 43} (2011) 1883
  [arXiv:1012.5069 [astro-ph.CO]].
\bibitem{ioav}  R.~O.~Ramos, M.~V.~d.~Santos and I.~Waga,
  ``Matter creation and cosmic acceleration,''
  Phys.\ Rev.\ D {\bf 89} (2014) 083524
  [arXiv:1404.2604 [astro-ph.CO]].
\bibitem{pan} S.~Pan and S.~Chakraborty,
  ``Will there be future deceleration? A study of particle creation mechanism in non-equilibrium thermodynamics,''
  arXiv:1404.3273 [gr-qc].
\bibitem{berezin} V.~Berezin,
  ``On the phenomenological description of particle creation and its influence on the space-time metrics,''
  arXiv:1404.3582 [gr-qc].
  \bibitem{Chakraborty:2014oya}
  S.~Chakraborty and S.~Saha,
  ``Inflation and late time acceleration: A unified prescription from the mechanism of particle creation,''
  arXiv:1404.6444 [gr-qc].
\bibitem{mccrea} W.H. McCrea, Relativity Theory and Creation of Matter,
Proc.Roy.Soc. of London, Series A, 206, 562 (1951)
\bibitem{oliver} O.~F.~Piattella, D.~C.~Rodrigues, Júl.~C.~Fabris and J.~é A.~de Freitas Pacheco,
  ``Evolution of the phase-space density and the Jeans scale for dark matter derivedfrom the Vlasov-Einstein equation,''
  JCAP {\bf 1311} (2013) 002
  [arXiv:1306.3578 [astro-ph.CO]].
\bibitem{planck} P.~A.~R.~Ade {\it et al.}  [Planck Collaboration],
  ``Planck 2013 results. XVI. Cosmological parameters,''
  arXiv:1303.5076 [astro-ph.CO].
\bibitem{gonzalez} A.~H.~Gonzalez, D.~Zaritsky and A.~I.~Zabludoff,
  ``A Census of Baryons in Galaxy Clusters and Groups,''
  Astrophys.\ J.\  {\bf 666} (2007) 147
  [arXiv:0705.1726 [astro-ph]].
\bibitem{ma} C.~-P.~Ma and E.~Bertschinger,
  ``Cosmological perturbation theory in the synchronous and conformal Newtonian gauges,''
  Astrophys.\ J.\  {\bf 455} (1995) 7
  [astro-ph/9506072].
\bibitem{hudson}
  M.~J.~Hudson and S.~J.~Turnbull,
  ``The growth rate of cosmic structure from peculiar velocities at low and high redshifts,''
  Astrophys.\ J.\  {\bf 751} (2013) L30
  [arXiv:1203.4814 [astro-ph.CO]].
\end{thebibliography}

\label{lastpage}

\end{document}